\documentclass{aa}  

\usepackage{graphicx}
\usepackage{txfonts}
\usepackage[colorlinks,linkcolor={blue},citecolor={blue}]{hyperref}

\begin{document} 
   
   \title{Theoretical wind clumping predictions from 2D LDI models of O-star winds at different metallicities}
     
  \titlerunning{Wind clumping and metallicity}
  
   \author{F.~A.~Driessen \and J.~O.~Sundqvist \and A.~Dagore}
   \institute{Institute of Astronomy, KU Leuven,
              Celestijnenlaan 200D, 3001 Leuven, Belgium\\
              \email{florian.driessen@kuleuven.be}
             }
             
   \date{Received ....; accepted ...}

 
  \abstract
   {Hot, massive (OB) stars experience strong line-driven stellar winds and mass loss. As the majority of efficient driving lines are metallic, the amount of wind driving and mass loss is dependent on the stellar metallicity $Z$. In addition, line-driven winds are intrinsically inhomogeneous and clumpy. However, to date, neither theoretical nor empirical studies of line-driven winds have investigated how such wind clumping may also depend on $Z$.}   
   {We theoretically investigated the degree of wind clumping due to the line-deshadowing instability (LDI) as a function of $Z$. }
   {We performed two-dimensional hydrodynamic simulations of the LDI with an assumed one-dimensional radiation line force for a grid of O-star wind models with fixed luminosity, but with different metal contents by varying the accumulative line strength $\bar{Q}$ describing the total ensemble of driving lines.}
   {We find that, for this fixed luminosity, the amount of wind clumping decreases with metallicity. The decrease is clearly seen in the statistical properties of our simulations, but is nonetheless rather weak; a simple power-law fit for the dependence of the clumping factor $f_{\rm cl} \equiv \langle \rho^2 \rangle/\langle \rho \rangle^2$ on metallicity yields $f_\mathrm{cl} \propto Z^{0.15 \pm 0.01}$. This implies that empirically derived power-law dependencies of mass-loss rate $\dot{M}$ on metallicity -- which were previously inferred from spectral diagnostics effectively depending on $\dot{M} \sqrt{f_{\rm cl}}$ but without having any constraints on $f_{\rm cl}(Z)$ -- should be only modestly altered by clumping. We expect that this prediction can be directly tested using new data from the Hubble Space Telescope Ultraviolet Legacy Library of Young Stars as Essential Standards (ULLYSES) project.}
   {}
   

   \maketitle
%

\section{Introduction}

The outflows and mass loss from hot, massive (OB) stars are predominantly caused by momentum transfer of photons scattering in spectral lines of ionised elements \citep[][hereafter CAK]{1975ApJ...195..157C}. This `line driving' is very efficient because the Doppler effect allows a spectral line to sweep out (deshadow) and interact with continuum photons  over a significant frequency range. Because the continuum photons interact with primarily metal atomic species, there is a direct dependence on the abundance and metallicity for both the wind driving and mass loss, which is confirmed by theoretical (steady-state) line-driven wind models \citep{2001A&A...369..574V,2008A&A...482..945G,2018A&A...612A..20K,2021A&A...648A..36B,2021MNRAS.504.2051V}, as well as by observations \citep{2007A&A...473..603M,2011ApJ...741L...8T,2012A&A...543A..85H,2017A&A...600A..81R,2022arXiv220207811M}.

However, line-driven winds are intrinsically unstable, because the deshadowing of wind ions from the Doppler effect leads to a runaway effect, or line-deshadowing instability \citep[LDI;][]{1979ApJ...231..514M,1980ApJ...241.1131C,1984ApJ...284..337O}. In numerical simulations of the LDI, the wind is disrupted due to strong reverse shocks that lead to an inhomogeneous wind consisting of overdense clumps and a sparse interclump medium \citep{1988ApJ...335..914O,2017MNRAS.469.3102F,2018A&A...611A..17S,2019A&A...631A.172D,2021A&A...648A..94L}. 

This wind clumping then also severely affects the radiative transfer modelling needed to derive empirical mass-loss rates from comparisons to observed spectra. Indeed, neglecting clumping may lead to empirical mass-loss rate estimates that differ by more than an order of magnitude for the same star, depending on which observed spectral diagnostics is considered \citep{2006ApJ...637.1025F}. Very recently, multi-diagnostic studies properly accounting for wind clumping (both optically thin and thick) were performed for a small set of O-supergiant stars in the Galaxy \citep{2021A&A...655A..67H}, enabling first systematically derived constraints on statistical wind clumping properties. Nonetheless, wind clumping in OB-stars outside the Galaxy is still poorly investigated and constrained, both theoretically and observationally (although see \citealt{2022arXiv220211080B}). This lack of constraints for clumping as a function of metallicity may then also indirectly affect the above-mentioned mass loss--metallicity relation for line-driven winds. For example, \citet{2007A&A...473..603M} performed a large sample study of O-stars from the Galaxy and the Large and Small Magellanic Cloud (LMC and SMC) to infer empirical mass-loss rates. Using primarily the clumping-sensitive optical H$\alpha$ line, these authors confirmed that LMC and SMC stars show lower mass-loss rates for a given stellar luminosity. However, \citet{2007A&A...473..603M} did not take into account wind clumping corrections in their mass loss--metallicity relation, meaning that the quantity they effectively really derived was $\dot{M} \sqrt{f_{\rm cl}} \propto Z^x$, with $x$ a power-law index and $f_{\rm cl}$ the (quantitatively unknown) clumping factor (as defined, e.g.~by Eq.~\eqref{Eq:fcl} of this paper).  With the advent of the Hubble Space Telescope Ultraviolet Legacy Library of Young Stars as Essential Standards (ULLYSES) project \citep{2020RNAAS...4..205R}, large amounts of high-quality spectra of extragalactic massive stars in the ultraviolet (UV) will be added to already existing optical data. Using similar techniques to those in \citet{2021A&A...655A..67H}, we expect that these observations will be able to provide for the first time quantitative empirical constraints on the metallicity dependence of wind clumping. As such, it becomes pertinent that this then can be adequately compared to corresponding theoretical predictions stemming from state-of-the-art modelling of the unsteady line-driven wind. To date, all such theoretical investigations of line-driven wind clumping have been performed for Galactic stars, meaning the expected dependence on metallicity remains mostly unexplored.  

In this paper, we investigate the dependence of wind clumping on metallicity from a theoretical point of view, using numerical radiation-hydrodynamic simulations of the LDI for O-stars at a fixed luminosity, but at different metallicities. The remainder of this paper is organised as follows: in Sect.~\ref{sec:hydro} we detail our hydrodynamic setup, after which we present our results in Sect.~\ref{sec:sims}. We discuss these in Sect.~\ref{sec:discussion} and conclude in Sect.~\ref{sec:conc}.

\section{Method}\label{sec:hydro}

We adopt the same numerical algorithms and grid setup as presented in \citet[][for specific details]{2021A&A...656A.131D} within the open-source astrophysical (magneto-)hydrodynamic code \textsc{mpi-amrvac} \citep{2018ApJS..234...30X,KEPPENS2021316}, except that we do not consider magnetic fields in the present work. This means we solve the equations of hydrodynamics on a two-dimensional (2D) spherical $(r,\theta)$ grid for density $\rho$ and velocity $\mathbf{v}$,
\begin{equation}
\frac{\partial \rho}{\partial t} + \nabla \cdot (\rho\mathbf{v}) = 0,
\end{equation}
\begin{equation}
\frac{\partial (\rho\mathbf{v})}{\partial t} + \nabla \cdot (\rho \mathbf{v} \mathbf{v} + p\mathbf{I}) = \mathbf{g}_\mathrm{eff} + \mathbf{g}_\mathrm{line},
\end{equation}
\begin{equation}
p = a^2\rho, \qquad a = \sqrt{k_{\rm B} T_\mathrm{wind}/m}\,.
\end{equation}

For the purpose of this study, we assume an isothermal wind at a temperature\footnote{We note that $T_{\rm eff}$ is actually slightly lower than $T_{\rm wind}$ using the parameters in Table \ref{tab:starparam}. However, this is of minor importance because in these isothermal models, the temperature dependence only manifests itself in the thermal pressure gradient of the momentum equation (which plays a minor role in setting the dynamics of the supersonic line-driven wind outflow).} $T_\mathrm{wind}\approx T_\mathrm{eff}$ with isotropic thermal pressure $p\mathbf{I}$ and isothermal sound speed $a$, which is a good approximation for the dense outflows from O-stars that undergo efficient radiative cooling \citep[but see][]{1995A&A...299..523F,2021A&A...648A..94L}. In the momentum equation, we take into account the effect of stellar gravity reduced by constant electron scattering $\Gamma_{\rm e}$, $\mathbf{g}_\mathrm{eff} = -GM_\star(1-\Gamma_{\rm e})/r^2$, and the radiation line force $\mathbf{g}_\mathrm{line}$. The adopted parameters in Table \ref{tab:starparam} also follow \citet{2021A&A...656A.131D} and are representative of a high-luminosity Galactic O-star. We vary the line-strength parametrisation $\bar{Q}$ to simulate different metallicity conditions (see Sect.~\ref{sec:sims} for details). A key difference with \citet{2021A&A...656A.131D} is that in this work, and for the first time in a 2D LDI wind model, we adopt a realistic line-strength cut-off $Q_{\rm max} = \bar{Q}$ (see Sect.~\ref{sec:sims}). Choosing a fixed luminosity $L_\star$ for our grid allows us to separate out another, potentially complicated relationship between clumping and $L_\star$, and thus to focus exclusively on the $Z$ dependence of interest here. The model luminosity $L_\star$ has been chosen such that it coincides well with that considered by \citet{2007A&A...473..603M} when they derived their empirical mass-loss metallicity relation for O-stars (neglecting the effects of clumping). 

\begin{table}
\caption{Overview of stellar and wind parameters.}             
{\small{

\centering    
\label{tab:starparam}
\begin{tabular}{l c c}       
\hline\hline                 
Name & Parameter & Value  \\  
\hline                        
   Stellar luminosity                    & $L_\star$ & $8\times 10^5\,L_\odot$   \\    
   Stellar mass                            & $M_\star$ & $50\,M_\odot$ \\
   Stellar radius                           & $R_\star$  & $20\,R_\odot$   \\
   Stellar wind temperature & $T_\mathrm{wind}$ & $40\,000$\,K    \\
   Eddington factor                     & $\Gamma_{\rm e}$  & 0.42 \\
   CAK exponent                         & $\alpha$ & $0.65$ \\
   Line-strength normalisation  & $\bar{Q}$ & $2000$\tablefootmark{a} \\
   Line-strength cut-off             & $Q_\mathrm{max}$ & $\bar{Q}$ \\
   Thermal to sound speed  ratio & $\varv_\mathrm{th}/a$ & 0.28 \\
   Isothermal sound speed         & $a$ & 23.3\,km\,s$^{-1}$\\
\hline                                   
\end{tabular}
\tablefoot{\tablefoottext{a}{This parameter is varied throughout our models while keeping all others fixed. Specifically, we take values of $\bar{Q}=Q_{\rm max}=$ 200, 400, 600, 800, 1000, 1200, 1400, 1600, 1800, and 2000.}}
}}
\end{table}

Following previous work on non-linear LDI simulations in a 2D hydrodynamic setting \citep{2003A&A...406L...1D,2021A&A...656A.131D}, we formulate the one-dimensional (1D) line force (consisting of direct and diffuse contributions) as $\mathbf{g}_\mathrm{line}=g_\mathrm{line}\,\mathbf{\hat{e}}_r = (g_\mathrm{dir} + g_\mathrm{diff})\,\mathbf{\hat{e}}_r$ in the radial direction for each co-latitudinal cell using the smooth source function approximation \citep[SSF;][]{1996ApJ...462..894O}. This avoids complex ray integrations from a bundle of rays with arbitrary directions. The adopted radial ray at each colatitude is then tilted with respect to the radial direction to mimic the finite extent of the stellar disc \citep{2013MNRAS.428.1837S} and to avoid that each point in the wind becomes a critical point (CAK). We refer the reader to \citet[][their Sect.~5]{1996ApJ...462..894O} and \citet[][their Eqs.~(9)$-$(14)]{2021A&A...656A.131D} for an extensive account of the implementation of $g_\mathrm{dir}, g_\mathrm{diff}$ into the hydrodynamics.

Inner and outer radial boundary conditions on the hydrodynamic variables are set in the same fashion as in \citet{2021A&A...656A.131D}, except that for the radiation field emerging from the star we do not assume a uniformly bright disc, but apply an Eddington limb-darkening law,
\begin{equation}
D(\mu,r) = \frac{1}{2} + \frac{3}{4}\mu' = \frac{1}{2} + \frac{3}{4}\sqrt{\frac{\mu^2 - \mu_\star^2}{1-\mu_\star^2}}, 
\end{equation}
with $\mu'=\cos \theta'$ the angle between a star-centred radius vector and a ray emanating from the stellar surface, $\mu$ the impact parameter, and $\mu_\star$ the dilution factor. This modifies at the lower radial boundary the direct radiation field $g_{\rm dir}$ contribution \citep[][for more details]{2013MNRAS.428.1837S,2019A&A...631A.172D}. To avoid numerical complications at the poles, all hydrodynamic variables at the (singular) polar axes are set using $\pi$-boundary conditions to connect both poles and allow flow from one side to the other without an artificial boundary.

As the radiation transport is only solved for in the radial direction (with a fiducial tilt, see above) this means that non-radial radiation effects are inhibited within our simulations. In particular, we are ignoring the line-drag effect that linear analyses suggest may damp lateral velocity fluctuations on scales below the lateral Sobolev length \citep{1990ApJ...349..274R,2020MNRAS.499.4282D}. Recent 2D radiation transport simulations of the LDI by \citet{2018A&A...611A..17S} (albeit on a spatially restricted grid) indeed show that typical lateral structures seem to be on the order of the Sobolev length ($\sim 0.01R_\star$ at $r \sim 2 R_\star$). This then motivates our choice here of a lateral grid, ensuring that a typical lateral Sobolev length (or arc length angle), $L_\theta /r \equiv \varv_\mathrm{th}/\varv_r \approx 0.5^\circ$, is resolved within our simulations.

The simulations start from a 1D, spherically symmetric, steady-state CAK wind that applies the Sobolev approximation in the computation of $g_{\rm line}$. The corresponding initial smooth density and radial velocity profiles are distributed over the meridional plane while setting lateral velocities to zero. From this initial state, we evolve the radiation-hydrodynamic equations using the SSF formalism for a total physical time of approximately 300\,ks (corresponds to about 30 wind dynamical timescales: $\tau_\mathrm{dyn} = R_\star/\varv_\infty \approx 10$\,ks).


\section{Simulation results}\label{sec:sims}

In this work, we do not rely on the original CAK line-force parametrisation but use instead the conceptually advantageous but equivalent $\bar{Q}$ parametrisation proposed by \citet{1995ApJ...454..410G}. For our purposes here, a key advantage of the line strength $\bar{Q}$ is its direct proportionality to metallicity $Z$ (\citealt{1995ApJ...454..410G}, his Eq.~(60); \citealt{2000A&AS..141...23P}, their Eq.~(81)),
\begin{equation}
\bar{Q} \propto Z.  
\end{equation}

A reference value is chosen such that the line-strength parameter takes the standard value $\bar{Q}=2000$ found for O-stars at solar metallicity $Z_\odot$ \citep{1995ApJ...454..410G,2000A&AS..141...23P}. In order to then investigate how LDI-generated structure depends on $Z$, we simply vary $\bar{Q}$ accordingly; that is, for example a simulation at $\bar{Q} = 1000$ approximately corresponds to a metallicity $Z/Z_\odot = 0.5$ (LMC) and one at $\bar{Q} = 400$ to $Z/Z_\odot = 0.2$ (SMC). 

A priori, the direct scaling of $\bar{Q}$ purely with $Z$ might appear surprising. However, the ensemble line strength $\bar{Q}$ is only dependent on the properties of the radiation field, $w_{\nu_i} = F_{\nu_i} \nu_i /F$, for flux $F$ and frequency of transition $\nu_i$, and the strength of a given single spectral line $q_i$ (\citealt{1995ApJ...454..410G}, his Eq.~(14); \citealt{2000A&AS..141...23P}, their Eq.~(37)),
\begin{equation}
\bar{Q} \equiv \sum_{i=1}^{\rm all\,lines} w_{\nu_i} q_i.
\end{equation}

The latter here depends on the atomic level occupation number density $n_l$ (neglecting stimulated emission), 
\begin{equation}
q_i \equiv \frac{ \sigma_{\rm cl} f n_l}{\sigma_{\rm Th} \nu_i n_{\rm e}}, 
\end{equation}
with $f$ the oscillator strength, $\sigma_{\rm cl} = \pi e^2/(m_{\rm e} c)$ the classical frequency-integrated line cross-section for an electron with charge $e$ and mass $m_{\rm e}$, $\sigma_{\rm Th}$ the Thomson cross-section, and $n_{\rm e}$ the electron number density. As the vast majority of driving elements are metals  in O-stars, it becomes clear from this definition that $q_i \propto n_l \propto A \propto Z$, that is $q_i$ (and thus $\bar{Q}$) directly relates to the global metallicity $Z$ via the elemental abundance $A$. On the other hand, for cooler A-stars ($T_{\rm eff} \sim 10\,000$\,K), contrary to the hotter O-stars studied in this work, line-driving from hydrogen lines might also be important, meaning that this simple linear relation between $\bar{Q}$ and $Z$ may not be completely valid in that regime \citep[][see discussion below their Eq.~(81)]{2000A&AS..141...23P}. Therefore, determining the exact dependence of the $\bar{Q}(Z)$ relation for any given effective temperature is a topic for future research.

\subsection{Model calibration and  dynamics of the structured wind}\label{sec:calib}

As we keep all stellar parameters except the metallicity fixed, our simulations should recover some standard scaling relations for the mass-loss rate as a function of metallicity (or more precisely as a function of our metallicity proxy $\bar{Q}$; see above). To probe the consistency of our wind models in this respect, we closely monitor the mass-loss rate in the different simulations. It is important to stress here that we do not predict mass-loss rates from our LDI wind models. On the contrary, we choose our line-force parameters $(\alpha, \bar{Q})$ such that the corresponding 1D relaxed CAK simulation provides a reasonable mass flux for a typical O-star at the model luminosity. Performing non-linear LDI simulations then results in a time-averaged mass-loss rate $\langle \dot{M} \rangle$ that should approximate this mass flux. 

\begin{figure}[t]
\centering
\includegraphics[width=\hsize]{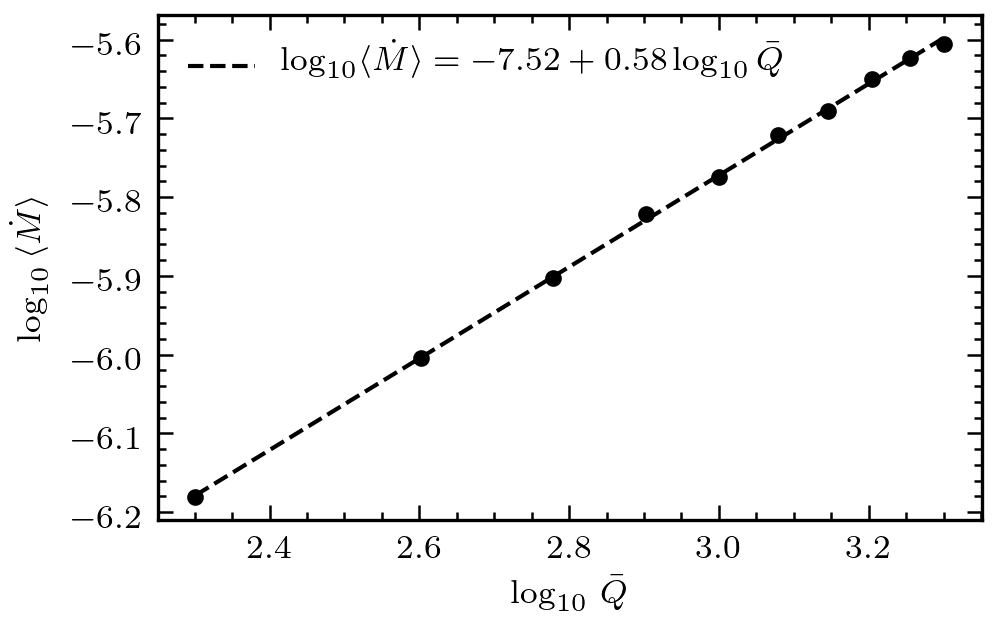}
\caption{Spatial and temporal averaged mass-loss rate $\langle \dot{M} \rangle$ as a function of line strength $\bar{Q}$ found from our simulations (dots). The dashed line is a power-law fit to the wind models.}
\label{fig:MdotQbar}
\end{figure}

Figure~\ref{fig:MdotQbar} shows the variation of $\langle \dot{M} \rangle$ with line strength. As expected, the mass-loss rate decreases with decreasing line strength because of the weaker wind. The decrease in mass loss between different line strengths can be compared to the theoretically expected reduction following analytic CAK theory. We take as reference the mass-loss rate $\dot{M}_{\bar{Q}_{2000}}$ for a Galactic star at solar metallicity, whereby:   
\begin{equation}
\dot{M}_{\bar{Q}} = \left( \frac{\bar{Q}}{\bar{Q}_{2000}} \right)^{(1-\alpha)/\alpha} \dot{M}_{\bar{Q}_{2000}}.
\end{equation}

As all parameters are fixed within our model, except for $\bar{Q}$, the mass-loss rate at a given metallicity and fixed $\alpha=0.65$ should scale as $\dot{M} \propto \bar{Q}^{0.53}$. When performing a power-law fit of the dependency of mass loss  on line strength in Fig.~\ref{fig:MdotQbar}, we indeed recover a power law with an index of $0.58$. Given the non-linear nature of our simulations, the variation from the analytical CAK value is well within the acceptable range. Moreover, our $\bar{Q} = 2000$ reference model yields $\log_{10} \langle \dot{M} \rangle = -5.61$, which also compares reasonably well with the detailed steady-state predictions by \citet{2021A&A...648A..36B} for a star of the same luminosity and $Z=Z_\odot$ ($\log_{10} \dot{M} = -5.75$). We do note that \citet{2021A&A...648A..36B} formally include millions of individual spectral lines to compute the radiative acceleration in their 1D stationary models. Within such detailed time-independent models, it takes typically $\approx 30$\,minutes to converge the non-local thermodynamic equilibrium radiative transfer and compute the line force for one hydrodynamical iteration (a converged wind model then typically needs $\approx 50$ such iterations). Following a similar approach within a time-dependent code is computationally unfeasible because it requires solving the formal integral of radiative transfer for every relevant spectral line at every radial grid point and at each hydrodynamic time-step. For example, a typical simulation presented here needs $\approx 1.5\times 10^5$ iterations to march towards $t_{\rm end}\approx 300$\,ks, which shows the impractical nature of such detailed line-transfer calculation for the radiative acceleration\footnote{Meaning a total computation time of approximately $4.5\times 10^6\,\text{min}= 3125\,\text{days}\approx 8.5\,\text{years}$ for a single hydrodynamic model.}. Instead, the line-ensemble parametrisation adapted here avoids such complications and statistically characterises the cumulative effect of all lines with an exponentially truncated power-law distribution and parameters $(\alpha, \bar{Q}, Q_\mathrm{max})$. As the obtained reference mass-loss rate agrees well with more detailed computations, we consider our wind models to be well-calibrated with respect to the chosen Galactic standard model of $\bar{Q}=2000$, such that our further predictions with respect to the wind clumping should be robust in a relative sense.

\begin{figure}[t]
\centering
\includegraphics[width=\hsize]{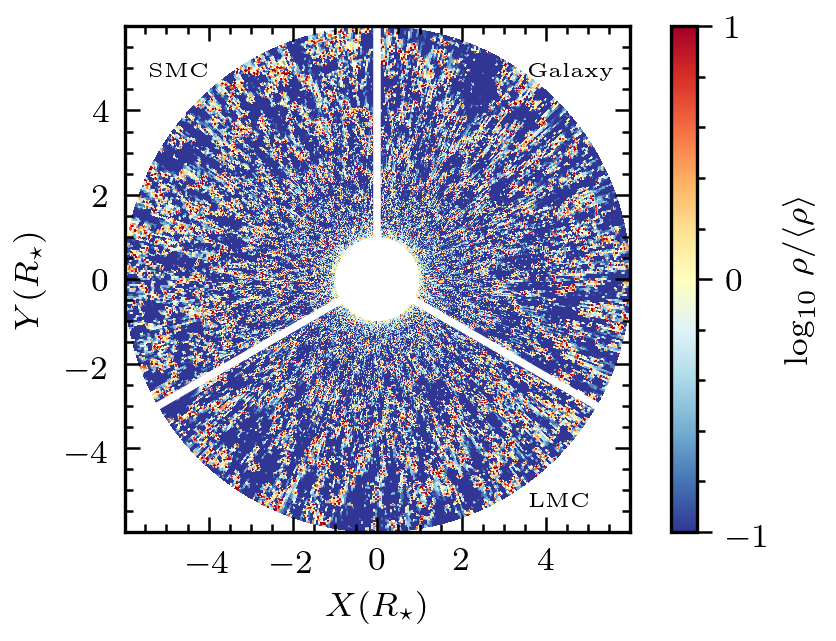}
\caption{Comparison of logarithmic density variations with respect to the mean wind density between a Galactic metallicity ($\bar{Q}=2000$), LMC metallicity ($\bar{Q}=1000$), and SMC metallicity ($\bar{Q}=400$) stellar wind.}
\label{fig:windcontours}
\end{figure}

Figure~\ref{fig:windcontours} shows the resulting wind density at a snapshot taken well after  (at 300\,ks) the LDI simulation has developed its characteristic structure consisting of high-density clumps and rarefied interclump gas. Specifically, we compare the Galactic wind model $\bar{Q}=2000$ ($Z \approx Z_\odot$) with a model mimicking the metallicity of the LMC, $\bar{Q}=1000$ ($Z=0.5Z_\odot$), and with one mimicking the SMC, $\bar{Q}=400$ ($Z=0.2Z_\odot$). All simulations display the characteristic LDI-generated features of a rather disrupted flow with quite slowly accelerated wind clumps that are separated by fast rarefied interclump gas. The lateral break-up of the LDI-generated structure is within these SSF models due to shearing motions together with Rayleigh--Taylor and thin-shell instabilities \citep{2003A&A...406L...1D} because the LDI itself only operates in the radial direction. Moreover, across this metallicity range, the wind clumps remain of similar size. However, this does not imply that the statistically computed overdensities of the clumps are the same (see Sect.~\ref{sec:fclspat}). Within the 1D SSF approximation, on what spatial scales LDI-generated clumps are expected to form is still unclear. Fragmentation might result in relatively large scales due to the lateral line-drag (not present in these models, \citealt{1990ApJ...349..274R,2020MNRAS.499.4282D}) or photospheric turbulent structures interacting with the LDI may also influence the clump spatial scales in the wind (see discussion in Sect.~\ref{sec:discussion}). Nonetheless, within the present model assumptions, the wind has a high degree of lateral incoherence whereby some clumpy structures only span a spatial scale close to the lateral cell size (i.e. they are on the order of the lateral Sobolev length). 

The effect of lateral break-up is likely enhanced here with our adaptation of a realistic line-strength cut-off: $Q_\mathrm{max}=\bar{Q}$ \citep{1995ApJ...454..410G}. This cut-off was originally introduced by \citet{1988ApJ...335..914O} to modify the CAK line-ensemble distribution and avoid an infinite radiative acceleration in the limit of optical depths approaching zero. However, in LDI simulations hitherto performed the cut-off has still been lowered to an artificial value of $Q_\mathrm{max} \approx 10^{-3}\bar{Q}$ to avoid severe numerical issues due to the overly steep acceleration of gas \citep{1995A&A...299..523F,2013MNRAS.428.1837S,2018A&A...611A..17S,2019A&A...631A.172D,2021A&A...656A.131D}. In our new 2D hydrodynamic simulations presented here, we are finally able to increase this cut-off to a physically realistic value, $Q_\mathrm{max} = \bar{Q}$. We speculate that the previous artificially low cut-off was inherent to the 1D LDI simulations that form the majority of the models present in the literature. Indeed, in these 1D models, gas parcels are forced to run into each other when the relative accelerations become too strong. The resulting strong shocks lead to strong density compressions, but also extremely rarefied gas that would get accelerated even more by virtue of its low optical depth, such that an artificially low line-strength cut-off is a necessity. These problems are alleviated in 2D models such as those presented here, where gas parcels are also able to flow past each other. As a result, some strongly accelerated rarefied gas is able to overtake an overdense clump instead of running into it and causing an even stronger shock. This then leads to an overall smoothing effect of the structures, which in turn means that we no longer have to cut off the line strength in order to prevent further growth. Hence, these new 2D simulations with $Q_{\rm max} = \bar{Q}$ should also represent a significant step forward regarding quantitative evaluation of clumping structures caused by the LDI (albeit still with the caveats discussed above concerning the lateral coherence scales).

\subsection{Spatial dependence of wind clumping}\label{sec:fclspat}

During our simulations we compute several statistical variables (i.e.~temporally averaged quantities) to characterise the wind. These statistical quantities are computed upon every iteration once the wind has sufficiently developed that it now displays its characteristic structure. Typically, this is reached within our simulation after a time of $t_\mathrm{stat}=200$\,ks, or about $20\tau_\mathrm{dyn}$, such that about 100\,ks in our simulations is devoted to computing the temporal averages.  

\begin{figure}[ht!]
\centering
\includegraphics[width=\hsize]{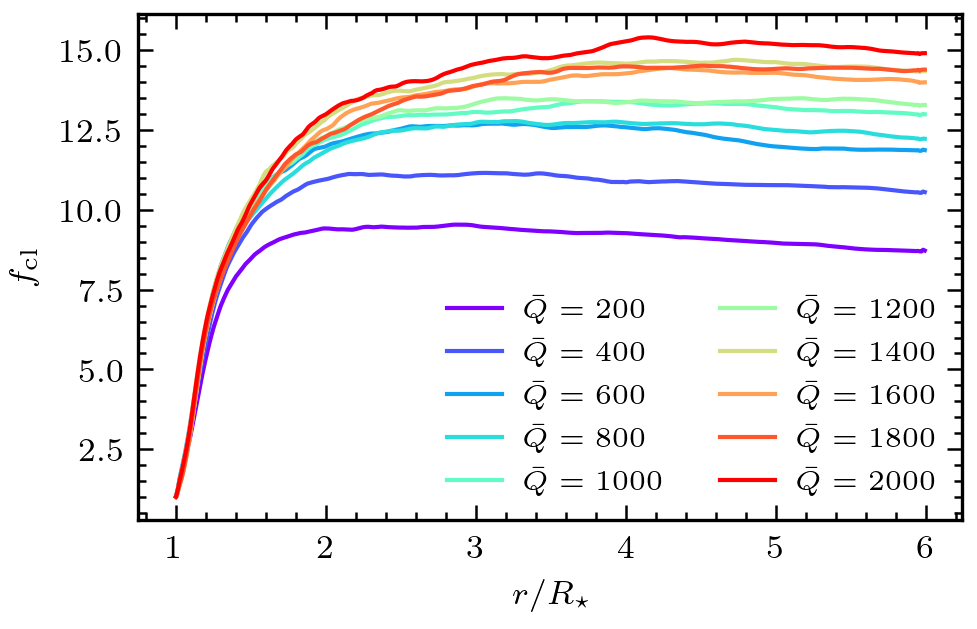}
\caption{Radial stratification (away from the star) of the wind clumping factor $f_\mathrm{cl}$ for each metallicity model.}
\label{fig:fclspatial}
\end{figure}

In this paper, we focus on the wind clumping factor, 
\begin{equation}
f_\mathrm{cl} \equiv \frac{\langle \rho^2 \rangle}{\langle \rho \rangle^2},
\label{Eq:fcl} 
\end{equation}
to quantify the amount of overdensity with respect to the mean wind density. As our models are 2D, we have a wind clumping factor in each point of the meridional plane. To that end, we compute from each model a corresponding lateral-averaged (1D) wind clumping profile in order to separate out the primary dependence on radius from the more stochastic lateral variation; the resulting $f_{\rm cl}(r)$ is displayed in Fig.~\ref{fig:fclspatial} for all simulations. The models confirm that, at lower metallicity, not only does the wind driving at a fixed luminosity become weaker, but so do the reverse shocks and density compressions that are caused by the LDI.

\begin{figure}[ht!]
\centering
\includegraphics[width=\hsize]{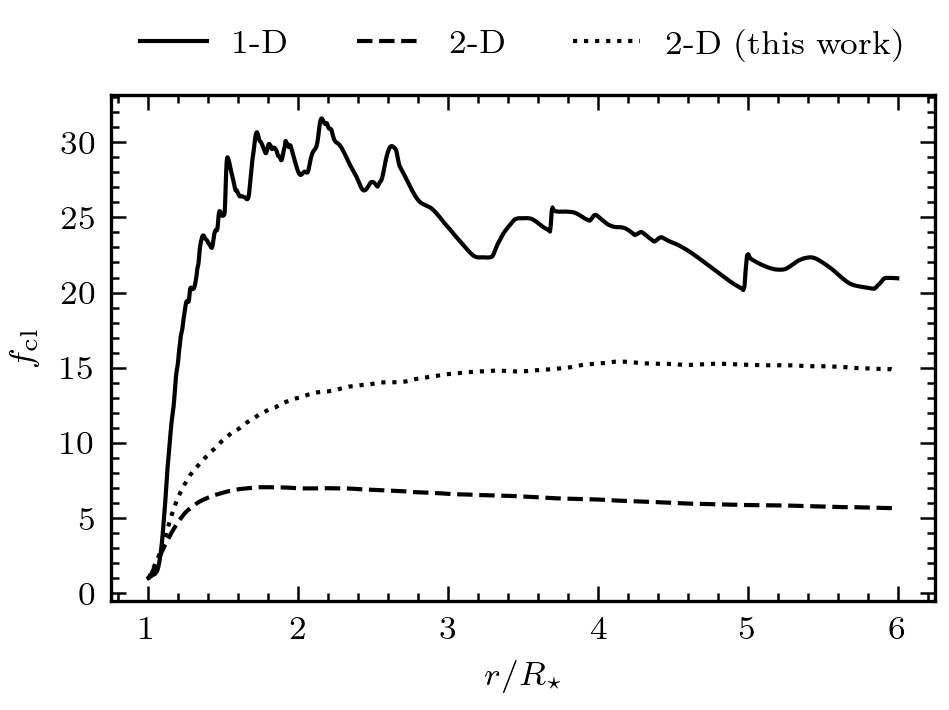}
\caption{Radial stratification (away from the star) of the wind clumping factor $f_\mathrm{cl}$ for the reference model $\bar{Q}=2000$ in different dimensions. The solid curve is a 1D simulation with $Q_{\rm max} = 10^{-3}\bar{Q}$ as found in the literature so far (but now computed with \textsc{mpi-amrvac}), while the dashed curve shows a comparable 2D simulation with $Q_{\rm max} = 10^{-3}\bar{Q}$. The dotted curve displays our novel 2D simulation with $Q_{\rm max}=\bar{Q}$. The stratification curves from the 2D models appear smoother because they are laterally averaged to retrieve a single radial stratification for comparison with a purely 1D model.}
\label{fig:fclcomp}
\end{figure}

We further find that, although structure formation generally starts already at the wind base, the growth in the inner wind is somewhat slower than in previous 1D models. 
To rule out that differences in wind clumping are due to the use of different codes (\textsc{mpi-amrvac} here; \textsc{vh-1} in previous studies\footnote{\texttt{http://wonka.physics.ncsu.edu/pub/VH-1/}}), we compare in Fig.~\ref{fig:fclcomp} the wind clumping stratification stemming from a standard 1D model with an artificial cut-off of $Q_{\rm max}=10^{-3}\bar{Q}$, a similar 2D model with $Q_{\rm max}=10^{-3}\bar{Q}$, and the 2D model with $Q_{\rm max}=\bar{Q}$ presented in this work, all for the reference line strength $\bar{Q}=2000$ and all computed with \textsc{mpi-amrvac}. We find a more modest degree of wind clumping for the 2D models in the region $r\leq 2R_\star$ as compared to 1D models \citep{2013MNRAS.428.1837S,2019A&A...631A.172D}. Nonetheless, our 1D model computed with \textsc{mpi-amrvac} agrees well with previous 1D models such that we cannot attribute the different wind clumping stratifications to the use of different codes. Specifically, in those 1D simulations of high-luminosity O-stars, the degree of wind clumping is found to strongly peak at $\approx 2 R_\star$ with $f_\mathrm{cl} \sim 20-30$ \citep{2013MNRAS.428.1837S,2019A&A...631A.172D} after which it starts to decrease significantly towards the outer wind. By contrast, from our new 2D simulations we find somewhat lower absolute values for the clumping factor as well as a less pronounced peak (see Figs. \ref{fig:fclspatial}, \ref{fig:fclcomp}).  Again we suspect that the same physics that governs the effect of the line-strength cut-off $Q_{\rm max}$ (see Sect.~\ref{sec:calib}) manifests itself also in setting the absolute degree of wind clumping. Indeed, the 2D nature of our models does not necessarily result in local gas-clump collisions that would lead to stronger shocks and enhanced density contrasts (thus higher wind clumping), and with an artificial cut-off to almost no degree of clumping. In a 1D model, this is a natural outcome such that 1D LDI wind models generally predict higher $f_\mathrm{cl}$ than corresponding 2D models \citep[see also][]{2003A&A...406L...1D,2018A&A...611A..17S}.

\subsection{Power-law dependence of wind clumping and metallicity}\label{sec:qbarZ}

The results shown above (Fig.~\ref{fig:fclspatial}) already suggest that the amount of wind clumping decreases somewhat with decreasing metallicity. Such an effect might indeed be expected, because a lower metallicity wind will undergo weaker wind driving that also results in weaker amplification of velocity fluctuations and subsequent shock formation. We here quantify to what degree wind clumping is dependent on metallicity by assuming a simple power-law relation between $f_{\rm cl}$ and $\bar{Q}$ ($\propto Z$). 

\begin{figure}[ht!]
\centering
\includegraphics[width=\hsize]{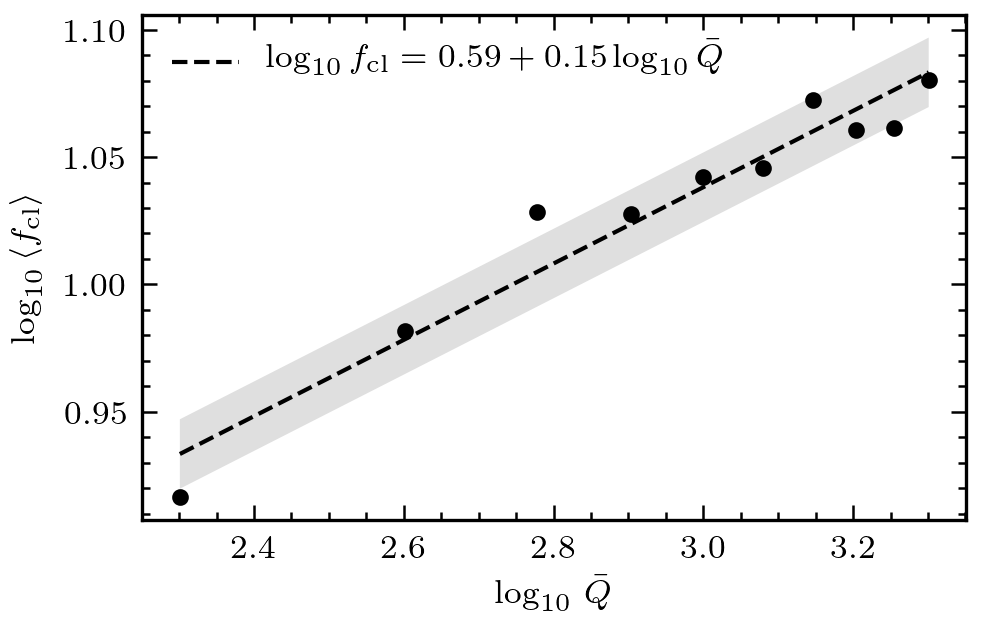}
\caption{Spatially- and temporally averaged wind clumping factor $\langle f_\mathrm{cl} \rangle$ as a function of line strength $\bar{Q}$ ($\propto Z$) found from our simulations (dots). The dashed line is a power-law fit to the wind models while the gray-shaded area gives the standard deviation $\sigma=0.01$ on the slope.}
\label{fig:fclQbar}
\end{figure}

Thus, we assume, 
\begin{equation}
f_\mathrm{cl} = a\bar{Q}^b \Leftrightarrow \log_{10} f_\mathrm{cl} = \log_{10} a + b\log_{10} \bar{Q},
\end{equation}
which represents a straight line in logarithmic space. The constant $a$ gives the intercept with the ordinate axis and the power-law index $b$ gives the slope of the straight line. To obtain one wind clumping estimate for the total wind, we compute from our 2D simulations the resulting lateral-averaged wind clumping (c.q.~Fig.~\ref{fig:fclspatial}) after which we further average this over the radial direction to get a characteristic $\langle f_{\rm cl} \rangle$ for the model. The results of this procedure are displayed in Fig.~\ref{fig:fclQbar} for each metallicity model, showing a clear increasing trend for $\langle f_{\rm cl} \rangle$ with increasing $\bar{Q}$. 

We apply a standard linear regression routine available from the Python \textsc{scipy} library \citep{2020SciPy-NMeth} to fit the simulation results to a power law with the above-mentioned form.  This yields: 
\begin{equation}
f_\mathrm{cl} \propto \bar{Q}^{0.15 \pm 0.01} \propto Z^{0.15 \pm 0.01}.
\end{equation}

The resulting fit therefore suggests a clear but still rather weak dependence of wind clumping on metallicity.

\section{Discussion} \label{sec:discussion} 

As mentioned in Sect. \ref{sec:hydro}, a key reason for our choice of characteristic model luminosity was the study by  \citet{2007A&A...473..603M}, who derived their empirical mass-loss--metallicity relation for stars in the same luminosity range. More specifically, \citet{2007A&A...473..603M} performed an extensive survey of O-type stars in the Galaxy, LMC, and SMC using the optical H$\alpha$ line, and derived an empirically calibrated mass-loss rate--metallicity relation:  
\begin{equation}
\dot{M} \propto Z^{0.83 \pm 0.16}.
\end{equation}

However, this relation was obtained by means of spectral fitting assuming smooth winds. As H$\alpha$ is a recombination line in O-stars, its emission will be very sensitive to the amount of clumping present in the wind. Namely, in an inhomogeneous wind, the mean opacity (per unit length) for H$\alpha$ scales as 
$\langle \chi \rangle \sim \langle \rho^2 \rangle \sim f_{\rm cl} \dot{M}^2$ for an assumed mass-loss rate $\dot{M} \sim \langle \rho \rangle$. That is, the scaling invariant is no longer proportional to $\dot{M}$ but rather to $\dot{M} \sqrt{f_{\rm cl}}$. As such, the empirical relation above should be modified according to: 
\begin{equation}
\dot{M} \sqrt{f_{\rm cl}} \propto Z^{0.83 \pm 0.16}.
\end{equation}

Inserting our theoretically derived wind clumping dependence on metallicity into this relation then finds a marginal correction to the mass loss--metallicity dependence:
\begin{equation}
\dot{M} \propto Z^{0.83}Z^{-0.15/2} = Z^{0.76}.
\end{equation}

Within uncertainties, this agrees fairly well with predictions made from steady-state line-driven wind models that do not rely on any underlying assumptions about the line-ensemble parameters when deriving the mass-loss rates \citep{2001A&A...369..574V,2018A&A...612A..20K,2021A&A...648A..36B}. As mentioned in the previous section, we further find somewhat lower degrees of clumping than corresponding 1D models in the inner wind, $r \la 2 R_\star$, and also do not observe any prominent peak for the clumping factor within our simulated domain. Using multi-wavelength diagnostic studies \citet{2006A&A...454..625P}, \citet{2011A&A...535A..32N}, and \citet{2022A&A...658A..61R} find that the inner wind of O-stars in the luminosity range considered here seem to be rather more clumped than the outermost radio-emitting wind. However, we note that the radio photosphere of such O-stars typically lies well beyond the outer boundary radius ($r/R_\star = 6$) of our 2D simulations. In Galactic 1D LDI models with much larger outer boundary radii, $f_\mathrm{cl} \approx 4$ has been found for the radio emission region at $r \ga 50R_\star$ \citep{2002A&A...381.1015R}; this is indeed lower than the $\langle f_{\rm cl} \rangle \approx 12$ derived here from our Galactic 2D simulation in the range $r/R_\star = 1-6$, which is in qualitative accordance with the observational findings mentioned above.  

Nevertheless, some recent studies focusing on multi-diagnostic spectral line fitting in the UV and optical found average clumping factors $f_{\rm cl} \approx 25$ for O-supergiants \citep[][]{2021A&A...655A..67H,2022arXiv220211080B}. Quantitatively, this is higher than the average values derived here, and may tentatively point towards some still missing physics in our simulations. In particular, considering that the 1D $\rightarrow$ 2D transition has led to overall lower degrees of clumping, it seems unlikely that, for example, a full three-dimensional simulation would reverse this trend. Alternatively, we might expect that adding a third dimension to the simulations would further decrease the quantitative levels of clumping, because then the high-density gas parcels will have yet another dimension to escape through that may lead to even more smoothing effects. 

In this respect, relaxing our assumption of a hydrostatic boundary at the stellar surface that connects to the wind might aid in inducing such higher levels of clumping also in the innermost wind. Specifically, for high-luminosity Galactic O-stars, the radiative acceleration might exceed gravity even in deep subsurface layers (at the so-called `iron-opacity bump’ at $T \approx 200$\,kK), and so it is possible that this could trigger a very turbulent lower atmosphere. Recently, \citet{2015ApJ...813...74J,2018Natur.561..498J} performed three-dimensional radiation-hydrodynamic simulations of massive-star envelopes and found that velocity and density fluctuations originating from the deep stellar layers are so vigorous that they can protrude into the stellar photosphere \citep{2020ApJ...902...67S}. Although these `blobs' can gain significant upward momentum from the subsurface opacity peaks, they typically stagnate and lead to `failed winds' falling back down below the surface, simply because these first simulations of subsurface layers relied on Rosseland mean opacities calibrated for static media and thus not including any line-driving effects. Recently, our group implemented suitable modules into \textsc{mpi-amrvac} for computing the relevant radiative energy and momentum terms \citep{2022A&A...657A..81M}, and further developed a hybrid opacity scheme able to simultaneously model the deep subsurface layers and a line-driven supersonic outflow in one unified approach \citep[][Poniatowksi et al, in prep.]{2021A&A...647A.151P}. With this, we will be able to investigate the effects of opacity-driven convective blobs on the line-driven wind structure and clumping properties in OB-type stars in future work. 

However, in this paper our primary interest has not been to obtain any absolute wind clumping factors, but rather to investigate their dependence on metallicity in a relative sense. As such, we consider our main simulation result to be robust, namely that the overall wind clumping decreases with metallicity for an O-star with fixed luminosity. In particular, effects stemming from subsurface layers will decrease with metallicity (primary opacity peak due to iron; see above), and so this should not change this overall qualitative behaviour.

Nonetheless, extensions to 3D and inclusion of a turbulent photosphere may slightly modify the quantitative behaviour of the clumping power law displayed in Fig.~\ref{fig:fclspatial}. However, in this respect, we note again that our 2D models are calibrated with respect to a standard Galactic reference model and so should be relatively robust. Proper consideration of the above-discussed effects may therefore change the wind clumping stratification in Fig.~\ref{fig:fclspatial}, but because we compute average wind clumping factors, any such difference in stratification will presumably be reflected in each model average. In other words, we expect at least the qualitative behaviour of wind clumping on metallicity to be relatively insensitive to further improvements of the underlying physics in radiation-hydrodynamic simulations of line-driven O-star winds. 

Finally, the models considered in this paper are all taken at a fixed reference luminosity in order to establish a first theoretical $f_{\rm cl}(Z)$ dependence. Following this approach allows a direct comparison with the empirical mass-loss study of \citet{2007A&A...473..603M}, who considered a similar reference luminosity, as discussed above. Nonetheless, it is likely that varying the stellar luminosity  would also affect the amount of wind clumping. In the linear analysis of \citet{2019A&A...631A.172D}, it was already found that the growth rate of the LDI (these authors did not consider metallicity effects in their simplified analysis) shows a complex dependence on stellar and wind parameters, and thus the amount of clumping in the non-linear regime may also depend on this indirectly. Yet, a potential luminosity dependence would add another free parameter to the current analysis such that it becomes difficult to disentangle the basic metallicity dependency we aim to investigate here. Whether the $f_{\rm cl}(Z)$ dependence derived above would also still hold for (significantly) different $L_\star$, or would show a more complex dependence on stellar parameters therefore remains an open question. 


\section{Conclusions and outlook}\label{sec:conc}

We present 2D radiation-hydrodynamic simulations of the LDI using the SSF approximation to evaluate the 1D line force \citep{1996ApJ...462..894O}. An important improvement of these 2D models is that we are able to, for the first time, assume a physically realistic line-strength cut-off when evaluating the radiation line force. 

In particular, we used our new simulations to investigate whether or not there is a dependence of wind clumping on metallicity for O-stars with a fixed luminosity. Our simulations suggest that   such a dependence indeed exists, albeit a rather weak one; fitting a power law to our models yields $f_\mathrm{cl} \propto Z^{0.15}$. Therefore, empirically inferred mass-loss rate--metallicity relations 
derived from diagnostics that probe $\dot{M} \sqrt{f_\mathrm{cl}}$ will only be modestly affected by our predicted $f_{\rm cl}(Z)$ relation; for example, the empirical $\dot{M} \propto Z^{0.83}$ derived by  \citet{2007A&A...473..603M} would be shifted to $\dot{M} \propto Z^{0.76}$. 

The $f_{\rm cl}(Z)$ relation derived in this work can be tested with upcoming spectra acquired as part of the ULLYSES project \citep{2020RNAAS...4..205R}. In particular, following \citet{2021A&A...655A..67H} and \citet{2022arXiv220211080B}, novel model atmosphere calculations combined with multi-diagnostic genetic algorithm fitting for large samples of stars should be able to put firm constraints on the empirical $f_{\rm cl}(Z)$ dependence. This would then allow us to assess whether our current 2D LDI simulations require further extensions and/or more input physics (see discussion in previous section), or if our predicted behaviour is robust against such further model improvements.

Additionally, such multi-diagnostic spectral fitting will also be able to establish better constraints on absolute wind clumping factors. This is of particular importance because our current 2D LDI wind models, contrary to previous 1D models, seem to quantitatively predict somewhat lower levels of $f_{\rm cl}$ than inferred from recent studies of Galactic O-supergiants \citep{2021A&A...655A..67H}. 

We further plan to re-examine our models by relaxing the assumption of a hydrostatic lower atmosphere in order to study the effect of opacity-driven convective and turbulent motions originating in the massive-star outer envelope protruding into the lower atmosphere \citep{2015ApJ...813...74J,2018Natur.561..498J}. The combined effect of such a time-dependent lower boundary and the interplay of the resulting `blobs' with the LDI can in principle be studied with our group's recently developed hybrid opacity-method \citep[][Poniatowski et al, in prep.]{2021A&A...647A.151P} and new radiation-hydrodynamic module in \textsc{mpi-amrvac} \citep{2022A&A...657A..81M}. This will then allow us to assess how such subsurface motions may also affect quantitative predictions of wind clumping, in particular in the inner wind. Furthermore, in published time-dependent line-driven wind models, the parameters $(\alpha, \bar{Q}, Q_\mathrm{max})$ (or different, but equivalent parametrisations) describing the driving from an ensemble of lines is taken as fixed both in space and time. To that end, here we focused our attention on a typical high-luminosity, early O-star for which the line-ensemble parameters are quite well constrained \citep{1995ApJ...454..410G,2008A&ARv..16..209P}. However, this fixing of the line-ensemble parameters for the full wind is a stringent condition that may have a significant impact on the wind dynamics. In future work, we plan to update and improve our LDI wind models (and clumping predictions) using tabulated line-driven wind parameters from detailed line-list computations that self-consistently compute $(\alpha, \bar{Q}, Q_\mathrm{max})$ based on the local hydrodynamic conditions of the wind (Poniatowski et al., in prep.). This would then also allow a larger systematic and self-consistent study of LDI-generated structures as a function of stellar parameters $(L_\star, M_\star, R_\star)$.

In summary, we consider the 2D models presented here to be the first step towards a more quantitative clumping analysis using numerical line-driven wind simulations at different metallicities. As discussed above, in that respect our study opens up many new avenues for the theoretical and observational study of structure formation in the winds of massive stars. 


\begin{acknowledgements}
FAD and JOS acknowledge support from the Odysseus program of the Belgian Research Foundation Flanders (FWO) under grant G0H9218N. We thank the anonymous referee for giving valuable comments that helped to clarify this work in several aspects.
\end{acknowledgements}

%
%

\bibliography{refs} 
\bibliographystyle{aa} 

\end{document}